\documentstyle[psfig,aabib]{l-aa}

\thesaurus{01(12.07.1; 12.04.1; 11.19.2; 11.17.4 B1600+434)}

\title{{Detection of a spiral lens galaxy and optical variability 
in the gravitational lens system B1600+434}\thanks{Based 
       on observations obtained with the {\it Nordic Optical 
       Telescope} (NOT), La Palma}}
\author{A.O. Jaunsen\inst{1,2} \and
J. Hjorth\inst{3}\thanks{{\it Present address}\/: NORDITA, Blegdamsvej 17,
DK--2100 K\o benhavn \O, Denmark (jens@nordita.dk)}}

\institute{
  Institute of Theoretical Astrophysics, P.O. Box 1029, Blindern, N--0315 Oslo, Norway (a.o.jaunsen@astro.uio.no)
  \and
  Nordic Optical Telescope, Ap.~474 St.~Cruz de La Palma, 
  E--38700 Canarias, Spain
  \and 
  Institute of Astronomy, Madingley Road, Cambridge CB3 0HA, UK}

\begin{document}
\date{\today}
\maketitle
\markboth{A.O. Jaunsen \& J. Hjorth: The gravitational lens system B1600+434}{}

\begin{abstract}
  The gravitationally lensed quasar B1600+434 ($z=1.61$, $m_V=21.6$)
  has been observed at the 2.56m Nordic Optical Telescope (NOT). In
  this Letter we report the discovery of an edge-on late-type galaxy
  located between the two lensed components (separation 1\farcs4),
  close to the fainter image.  The galaxy photometry indicates that
  its redshift is approximately $0.4$.  We detect a large colour
  difference between the two images due to significant obscuration of
  the faint image. The estimated amount of absorption as a function of
  colour indicates that the extinction may be due to dust in the
  lensing galaxy.  We also present evidence of flux variability in
  B1600+434 with a detected change of $\sim$0.25 mag in one year. The
  theoretically expected time delay is of the order of one month and
  so the system may be an interesting object for determining the
  Hubble constant.
\end{abstract}


\section{Introduction}

As part of the CLASS radio survey \cite*{Jackson1995} reported the
discovery of a new gravitational lens (GL) candidate, B1600+434,
with an image separation of $\theta = 1\farcs4$ and a flux ratio of
$I_A/I_B = 1.3 \pm 0.04$ at 8.4 GHz.  Traditionally, gravitationally
lensed images are expected to have identical colours due to the
achromatic nature of gravitational lensing.  In this Letter we report
our findings from optical multicolour observations of this system
obtained at three epochs which reveal a large colour difference
between the two components.

\section{Observations and reductions}
\label{sec:obsred}

\begin{figure}[t]
  \psfig{file=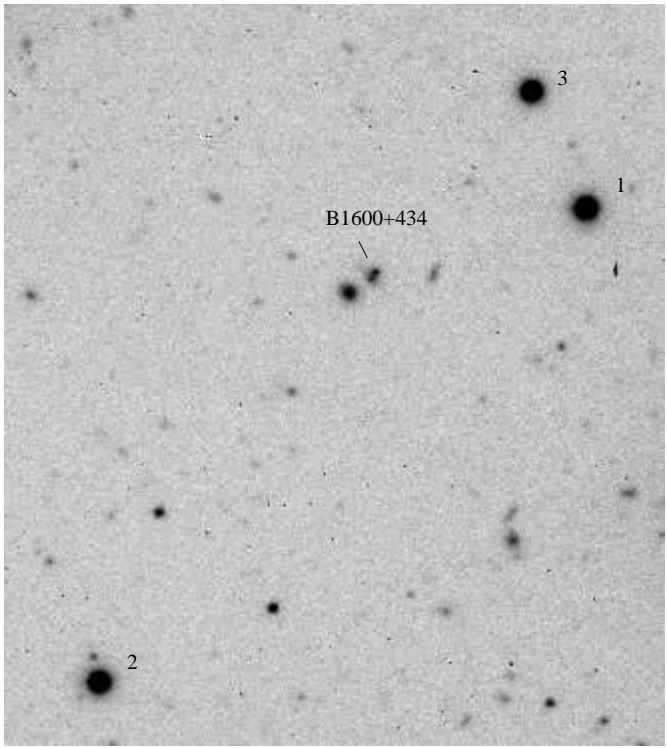,width=8.8cm} 
  \caption{Combined ($73 \times 83$ arcsec) image of the B1600+434 field. North           is up and east is to the left}
\label{fig:findchart}
\end{figure}

The observing log is given in Table~\ref{tab:obslog}. For the
observations we used the CCD cameras BroCam 1 (epochs 1 and 2) and
BroCam 2 (epoch 3). BroCam 1 is equipped with a thinned backside
illuminated TEK1024 CCD with a gain of $q = 1.7 {\rm e}^-\,{\rm
ADU}^{-1}$, a readout noise of $RON = 6.5 {\rm e}^-$ and a pixel scale
of $0\farcs176~{\rm pixel}^{-1}$.  BroCam 2 applied a thinned Loral
2048 CCD with $q = 1.27 {\rm e}^-\,{\rm ADU}^{-1}$, $RON = 6.22 {\rm
e}^-$ and a pixel scale of $0\farcs11~{\rm pixel}^{-1}$.  A finding
chart of the object and stars in the field is shown in
Figure~\ref{fig:findchart}, including the three stars used as internal
calibration and point-spread function (PSF) stars. In what follows we
label the Northern image of B1600+434 as image A and the fainter
Southern image as image B (cf.~Fig.~\ref{fig:lensgal}).
\vspace{-0.5cm}
\begin{table}
\begin{flushleft}
\caption{Dates and combined exposure times}
\label{tab:obslog}
\begin{tabular}{llllll}
  \hline
  Epoch & Date & B & V & R & I \\
  & & sec & sec & sec & sec \\ \hline
  1 & July 23--24 1995 & 1200 & 800 & 600 & 1200 \\
  2 & September 10--12 1995 & 1200 & 2300 & 600 & 1200 \\
  3 & June 7--10 1996 & 1800 & 2800 & 900 & 1200 \\ \hline
\end{tabular}
\end{flushleft}
\end{table}
\vspace{-0.5cm}

The CCD frames were overscan subtracted, zero-image subtracted and
flat fielded.  Each science frame was manually cleaned for cosmic rays
in the vicinity of stars and objects used in the analysis.

Standard stars in M5, M92 and a few Landolt fields were observed on
June 7 1996 (epoch 3) under photometric conditions and the three PSF
stars could thereby be calibrated to the Johnson-Kron-Cousins system.
This was achieved by using calibrated photometry from
\cite*{Sandquist1996} (M5), \cite*{Davis} (M92), \cite*{Grundahl} (M5,
M92) and \cite*{Landolt1992}.
The transformation equations were fitted to about 80 stars to derive
colour and extinction terms. The calibrated standard star magnitudes
could be reproduced to an rms of $0.02-0.03$ mag.  The extinction
coefficients determined at the Carlsberg-Automated-Meridian-Circle
(CAMC) for the La Palma site are in good agreement with the values
deduced from our observations.

\section{PSF photometry}
\label{sec:psfphot}

Using DAOPHOT II (\cite{Stetson1987}) we determined an analytical PSF
with a residual lookup-table from the three bright stars in the field.
The PSF photometry was carried out using ALLSTAR. A multiple fit of
the PSF to the QSO pair was constrained by fixing the position of B
relative to A using the radio positions. A background value similar to
the immediate surroundings was used. This yielded magnitudes of the
QSO images relative to the PSF stars.  The photometric zero-point was
determined from aperture photometry of the PSF stars.  The calibrated
magnitudes for A, B and the PSF stars are given in
Table~\ref{tab:magnBA} for each colour band in epoch~1, along with the
average measured FWHM values of the three stars.
\begin{table}
\begin{flushleft}
\caption{Photometry (mag) of image A, B and stars (epoch~1)}
\label{tab:magnBA}
\begin{tabular}{llllllll}
\hline
 & \hspace{-2.5mm}\mbox{FWHM} & A & B & B$-$A & 1 & 2 & 3 \\ \hline
B & 1\farcs21 & 22.12 & 23.82 & 1.70 & 18.97 & 18.74 & 18.73 \\
V & 1\farcs01 & 21.92 & 23.32 & 1.40 & 17.69 & 17.75 & 17.87 \\
R & 0\farcs60 & 21.46 & 22.34 & 0.88 & 16.81 & 17.12 & 17.29 \\
I & 0\farcs77 & 20.92 & 21.43 & 0.51 & 16.09 & 16.63 & 16.84 \\ \hline
\end{tabular}
\end{flushleft}
\end{table}

\section{Lensing galaxy}
\label{sec:lensgal}

\begin{figure}
\centerline{\psfig{file=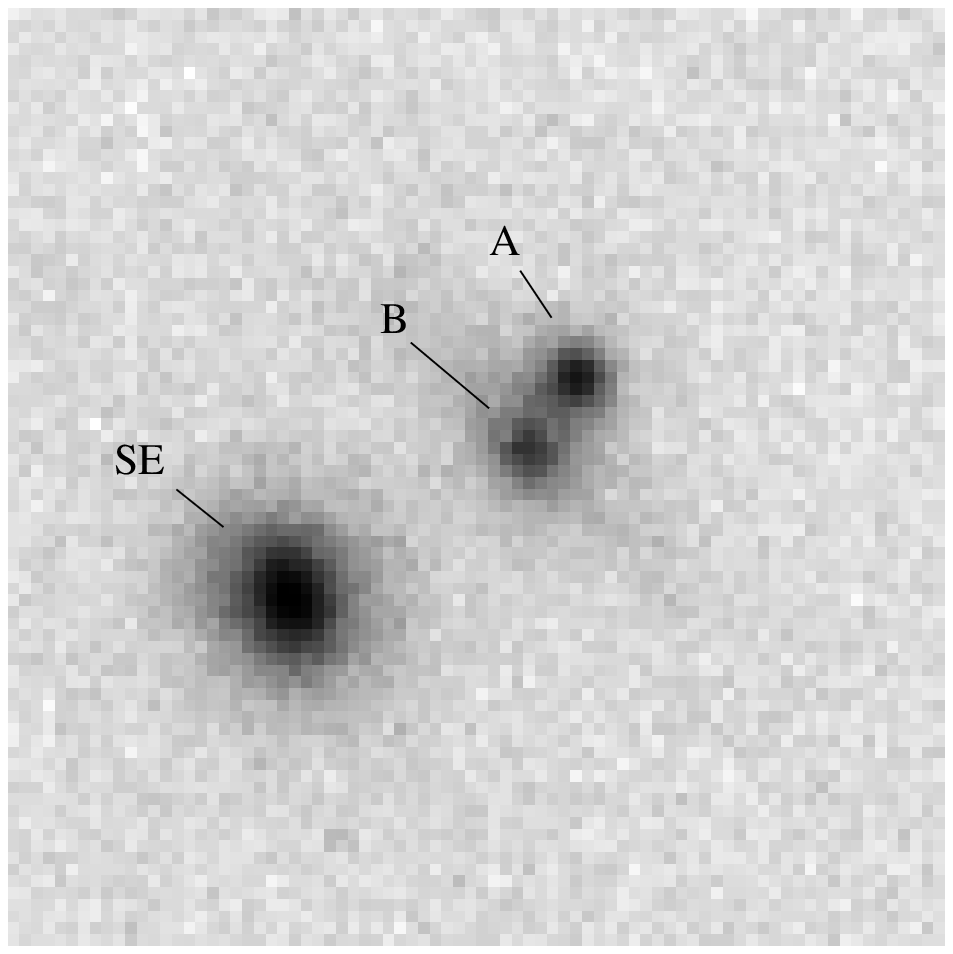,width=44mm}\mbox{\hspace{0.5mm}}\psfig{file=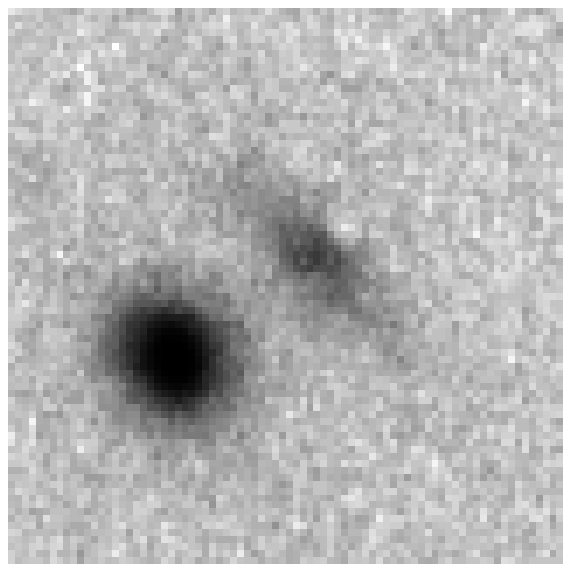,width=44mm}}
\caption{Combined V, R and I image of B1600+434 showing the A and B
  components (left). The same field after image A and B have been
  subtracted reveals the morphology of the lensing galaxy more clearly
  (right).  The early-type SE galaxy is also seen in the images.  The
  frames measure 14 arcsec on each side.  North is up and east is to
  the left}
\label{fig:lensgal}
\end{figure}

After subtracting the scaled PSFs from image A and B we are able to
study the residual lens galaxy located very close to image B (see
Fig.~\ref{fig:lensgal}).  We estimate the lens to have a major axis $a
= 4\farcs1 \pm 0\farcs2$, an axis ratio $a/b = 2.4 \pm 0.2$ and a
position angle PA~$=46\degr~\pm~3\degr$. These values were measured
from an isophote $1.2 \sigma$ above the sky using SExtractor
(\cite{Bertin1996}).  Astrometry of the lens, the South-East (SE)
galaxy and QSO images relative to image A (RA$_{2000}:
16^h01^m40\fs5$, Dec$_{2000}: +43\degr16\arcmin47\farcs0$) is given in
Table~\ref{tab:astrometry}.  We apply an elongated aperture with a
given major and minor axis to measure the intensity of the lensing
galaxy.  The sky is estimated from an outer elongated annulus
extending from 1.5 to 3.0 times the major and minor axis. To prevent
neighbouring galaxies from contaminating the sky estimate, overlapping
apertures of such objects were masked out.  We also measured the two
galaxies to the South-East (SE) and West (W) of B1600+434. To correct
for neighbouring stars or galaxies surrounding PSF stars 2 and 3 we
measured these stars with a small aperture radius and added aperture
corrections based on growth curve plots (\cite{Grundahl1995}) of star
1 (see Fig.~\ref{fig:findchart}) which had a nicely converging growth
curve in all images.  The photometric results are given in
Table~\ref{tab:lensgal}.
\vspace{-0.5cm}
\begin{table}
\begin{flushleft}
\caption{Astrometry of lens, SE galaxy and QSO images}
\label{tab:astrometry}
\begin{tabular}{lll}
  \hline & $\Delta \alpha$ & $\Delta \delta$ \\ \hline
  A & $0.0$ & $0.0$ \\
  B & $0\farcs72$ & $1\farcs20$ \\
  Lens & $0\farcs74 \pm 0\farcs10$ & $0\farcs85 \pm 0\farcs10$ \\
  SE & $4\farcs33 \pm 0\farcs05$ & $3\farcs27 \pm 0\farcs05$ \\ \hline
\end{tabular}
\end{flushleft}
\end{table}
\vspace{-1.5cm}
\begin{table}
\begin{flushleft}
\caption{Photometry (mag) of lensing galaxy}
\label{tab:lensgal}
\begin{tabular}{lllll}
\hline
 & B & V & R & I \\ \hline
Lens & $23.6$ & $22.0$ & $21.1$ & $20.3$ \\
SE & $22.0$ & $20.6$ & $19.9$ & $19.2$ \\
W & $23.9$ & $22.3$ & $21.7$ & $21.0$ \\ \hline
\end{tabular}
\end{flushleft}
\end{table}
\vspace{-0.5cm}

Using the $B-V$, $V-R$ and $R-I$ colour indices we estimated the
photometric redshifts of the three galaxies. These colours were fitted
(by least squares) to the template colours of various redshifts and
galaxy types tabulated in \cite*{Fukugita1995}.  On the basis of the
residual image and the measured galaxy properties we assume that the
lensing galaxy is a spiral galaxy. The photometry indicates
approximate redshifts of $z_l \sim 0.4$ for the lens, $0.1$ for the SE
galaxy (E/S0) and $0.2$ for the W galaxy (spiral).  The estimated
uncertainty in these redshifts is $0.1$.

\section{Galaxy extinction/absorption}
\label{sec:galabs}

A particularly interesting property of B1600+434 is the strong
apparent colour difference between the two images (see
Table~\ref{tab:magnBA}).
Assuming that the true intensity ratio of A and B is that measured in
the radio (\cite{Jackson1995}), we have
\begin{equation} 
\frac{I_A}{I_B}\bigg\vert_{\rm true} =
\frac{I_A}{I_B}\bigg\vert_{\rm radio} = 1.3.  
\end{equation} 
If in addition we assume that A is not affected by reddening, i.e.,
$I_{A_{\lambda}} = I_{A_{\rm true}}$ and $I_{B_{\lambda}} =
k_{\lambda}^{-1} I_{B_{\rm true}}$, the degree of absorption in the
galaxy is given by 
\begin{equation} 
k_{\lambda} = \frac{I_{A_{\lambda}} / I_{B_{\lambda}}}{I_{A_{\rm radio}} / I_{B_{\rm radio}}}, 
\label{eq:kabs} 
\end{equation} 
at restframe wavelength $\lambda$.  In Table~\ref{tab:rfabs} we give
the observed colour band, the central wavelength (CWL) bandpass, the
restframe CWL assuming $z=0.4$, the A:B intensity ratio and the
estimated absorption values.
\vspace{-0.5cm}
\begin{table}
\begin{flushleft} 
\caption{Lensing galaxy rest-frame absorption values} 
\label{tab:rfabs} 
\begin{tabular}{lllll} \hline
Filter & $CWL$ & $CWL_{z=0.4}$ & $I_A/I_B$ & $k_{\lambda}$ \\ \hline 
B & 4448 & 3177 & 4.79 & 3.68 \\ 
V & 5505 & 3932 & 3.63 & 2.79 \\ 
R & 6588 & 4706 & 2.25 & 1.73 \\ 
I & 8060 & 5757 & 1.60 & 1.23 \\ \hline
\end{tabular}
\end{flushleft}
\end{table}
\vspace{-0.5cm}
The reddening of image B is most likely due to dust obscuration in the
lensing galaxy.  The extinction due to dust increases roughly as
$k_\lambda\propto 1/\lambda$ (\cite{Mathis1990}). The R and I band
$k_\lambda$ estimates follow this trend, while the B and V gradients
are somewhat higher than expected.  This discrepancy is probably
partly due to errors in our PSF photometry of the B and V colours
because of the worse seeing in these images. The values do, however,
roughly agree with the expected trend for dust obscuration of image B.
Note that \cite*{Nadau1991} estimated the first differential reddening
of a gravitational lens system (2237+030) and found good agreement
with extinction due to dust in the Galaxy.

\section{Variability}
\label{sec:var}
To measure the possible effects of variability of the QSO we have
performed aperture photometry on the combined light of the lensed
images (A+B), the lensing galaxy and the SE galaxy.  This method was
chosen because the seeing was worse than one arcsec in several images
making the PSF subtraction uncertain for variability studies.  As a
check on our results we also used an aperture covering only the A+B
images and the lens and found similar results, but with larger
uncertainties (especially in the B band). To allow an optimal analysis
of the three epochs, the images were magnified, shifted and rotated to
the same coordinate system. A single circular aperture was centred
between the SE and lens galaxies.  Due to low signal-to-noise, errors
in the sky estimate and contamination by neighbouring objects, the
growth curve of the aperture did not converge at large radii. To
overcome this problem we reduced the aperture radius to a minimum of
$5\farcs0$, while still including most of the light. The flux outside
this radius was included by extrapolating the growth curves and
determining an aperture correction for each image. The sky was
estimated from a sky annulus with a radius of $8\farcs8$ extending to
$9\farcs9$.  The measured total magnitude of A+B+the lens+the SE
galaxy is listed relative to epoch 1 in Table~\ref{tab:var}.
\vspace{-0.5cm}
\begin{table}
\begin{flushleft}
\caption{Variability of A+B+lens+SE in the B, V, R and I bands relative to epoch 1}
\label{tab:var}
\begin{tabular}{lllll}
\hline
 & B & V & R & I \\ \hline 
$\Delta_{2-1}$ & $0.00 \pm 0.10$ & $0.11 \pm 0.02$ & $0.10 \pm 0.02$ & $0.18 \pm 0.02$ \\ 
$\Delta_{3-1}$ & $0.30 \pm 0.10$ & $0.25 \pm 0.03$ & $0.15 \pm 0.02$ & $0.21 \pm 0.02$ \\  \hline
\end{tabular}
\end{flushleft}
\end{table}
\vspace{-0.5cm}
The uncertainty in the B band magnitudes is approximately 4--5 times
greater than those in the other filters.  Apart from this filter the
remaining bands show a consistent variation. Due to the uncertainties
in the measurements it is difficult to identify whether the detected
variation is due to an intrinsic flux variation of the QSO or a
micro-lensing event. \cite*{Jackson1995} indicated flux variability in
their radio data and this is clearly supported by our optical data.

\section{Lens mass and time-delay}
\label{sec:massdt}

\cite*{Borgeest1984} introduced a radially symmetric galaxy model to
give rough estimates for the mass and time-delay in the QSO 0957+561
system.  We apply their model to estimate the lensing mass and
time-delay between image A and B in B1600+434. The lensing galaxy is
very close to the position of the B component and we found that the
separation between image A or B and the lens along the vector
connecting A and B is $\theta_A = 1\farcs11$ and $\theta_B=0\farcs29$.
The cosmological correction factor, T, is found in \cite*{Kayser1983}
for the lens and source redshifts and a flat Friedmann universe model
($q_0 = 0.5$ and $H_0= 100 h\, {\rm km\,s}^{-1}\,{\rm Mpc}^{-1}$ is
used throughout this Letter).  We assume there is no additional shear
from a cluster or other galaxies ($\gamma = 0$) and that the lens has
an isothermal mass distribution ($\kappa = 1$).  Inserting the
observational values into equation (15b) of \cite*{Borgeest1984} gives
a time-delay of $24\ h^{-1}$ days.  The enclosed mass inside the
angular radius $\theta/2$ (the Einstein radius) is found using their
equation (11) and gives $6.8 \ h^{-1} \ 10^{10}$ M$_\odot.$

As the lensing galaxy of this system is relatively bright it is
straightforward to estimate the mass-to-light ratio. We measure the
galaxy luminosity within a radius of
$\theta/2$~\mbox{($=0\farcs7=1.9$~kpc)} from the galaxy center and
compute the rest-frame absolute luminosity.  The rest-frame luminosity
of the galaxy is estimated using the method described in
\cite*{vanDokkum1996}; note that there is a sign error in their
definition of the conversion constants $c_i$. From the observed V, R
and I colours (given in Table~\ref{tab:lensgal}) we find the
rest-frame B and V magnitudes at $z=0.4$ to be $m_B = 23.6$ and $m_V =
22.8$. Assuming $h=0.7$ the corresponding absolute magnitudes are $M_B
= -17.7$ and $M_V = -18.6$.  This gives mass-to-light ratios of $51$
and $43$ in solar units for the rest-frame B and V bands,
respectively.  These relatively high M/L estimates are, however,
affected by various effects. Edge-on spiral galaxies commonly have
dust lanes which significantly obscure the luminous matter in the
galaxy (\cite{Barnaby1995}). Furthermore, the seeing smears out the
light inside the restricted aperture used and the PSF photometry
unavoidably introduces errors with a tendency for oversubtraction. Our
mass-to-light ratios should consequently be taken as upper limits.
Infrared observations of B1600+434 with HST would offer an optimal
measurement of the true lens galaxy M/L.

\section{Conclusions}
\label{sec:conclusions}

We find that a $V=22.0$ nearly edge-on spiral galaxy acts as the lens
in B1600+434 and causes a significant reddening of image B, due to
dust obscuration.  The reddening of image B is unusually strong and
may in general have implications on lensing statistics, because most
optically discovered GL candidates are chosen by their colour
similarity (e.g.~\cite{Jaunsen95}). Small deviations in the relative
magnitude for a given filter is usually allowed, but a difference of
the order of one magnitude or more will probably result in that object
being discarded as a GL candidate.  The failure to detect all GL
candidates in a sample will introduce systematic effects which affect
the results of the statistical analysis. \cite*{Kochanek1991}
discussed various bias effects in optical gravitational lens surveys
and concluded that 10--30 \% of all GL systems are lost already in the
process of identifying the QSOs. However, this is a rough estimate
which depends on the selection criteria of the QSO sample.  Indeed, in
view of the results presented here, the number of unidentified GL
systems in optical QSO/GL surveys is difficult to estimate.

If most of the GL systems discovered so far are produced by early-type
ellipticals this could be explained by the fact that optically thick
spiral galaxies tend to redden one of the (two) lensed images, thereby
causing many spiral lensed systems from being detected in optical
surveys.  Radio selected samples are much less affected by such
selection effects and may consequently reveal more spiral lenses. In
fact B1600+434 is the second radio-selected spiral lens galaxy
discovered at NOT by our group, the first being B0218+357
(\cite{Grundahl1995}) and both systems show signs of significant
extinction.

Our variability analysis indicates a variation of about 0.25 mag\,
year$^{-1}$. The lens system B1600+434 may therefore be an interesting
candidate for determining the time-delay and $H_0$. We estimate the
theoretically expected time-delay to be of the order of one month. It
should be noted, though, that the shear effects of the SE galaxy may
affect the lensing and should probably be accounted for in a realistic
model of the lens potential.

\begin{acknowledgements}
AOJ was supported by the Norwegian Research Council (NFR). JH
acknowledges financial support from the Danish Natural Science
Research Council (SNF).
\end{acknowledgements}

\end{document}